\documentstyle[aps]{revtex}
\begin{document}
\draft
\twocolumn
\title{Reply on the ``Comment on  
`Loss-error compensation in quantum-state measurements' ''}
\author{T. Kiss$^a$, U. Herzog$^b$, and U. Leonhardt$^c$}
\address{$^a$ Research Laboratory for Crystal Physics,
            Hungarian Academy of Sciences,
            P.O. Box 132, H-1502 Budapest, Hungary}
\address{$^b$ Arbeitsgruppe ``Nichtklassische Strahlung'',
Institut f\"ur Physik der Humboldt Universit\"at zu Berlin,
Rudower Chaussee 5, 12484 Berlin, Germany }
\address{$^c$Abteilung f\"ur Quantenphysik,
         Universit\"at Ulm, 89069 Ulm, Germany}
\maketitle
\begin{abstract}
The authors of the preceding Comment
[G. M. D'Ariano and C. Macchiavello 
Phys. Rev. A (preceding comment), quant-ph/9701009]
tried to reestablish a $0.5$ efficiency bound
for loss compensation in optical homodyne tomography.
In our reply we demonstrate that
neither does such a rigorous bound exist nor is the bound required 
for ruling out the state reconstruction of an individual system 
[G. M. D'Ariano and H. P. Yuen,
Phys. Rev. Lett. {\bf 76}, 2832 (1996)].
\end{abstract}
\date{today}
\pacs{03.65.Bz, 42.50.Dv}

There is little doubt that the compensation of detection losses is a 
numerically delicate procedure that is extremely sensitive to experimental
inaccuracies. We have shown \cite{KHL,Herzog}, however, that no clear 
efficiency bound exists beyond that loss compensation is impossible,
in contrast 
to a statement in an earlier paper \cite{DLP}. In the preceding Comment
\cite{DM} D'Ariano and Macchiavello tried to reestablish a $0.5$ bound
for the overall efficiency $\eta$.

In this Reply we show that their analysis is incomplete and that still no 
in-principle bound exists. Furthermore we point out
that the existence of such a compensation bound does not follow  
directly from the arguments given in Ref. \cite{DY}
where the impossibility of 
measuring the state of an individual quantum system was proven. 

What is the problem? We have shown \cite{KHL} 
that a generalized Bernoulli transformation describes the influence of 
detection losses on the density matrix $\langle m | \hat \rho | n \rangle$
that is reconstructed from measured data obtained e.g. from optical homodyne 
tomography \cite{OHT}. We have treated the loss process separately from the 
particular detection scheme. The inversion of the Bernoulli transformation 
\cite{KHL} produces the unperturbed density matrix of the signal
\begin{equation}
\langle m | \hat \rho_{sig}| n \rangle =
\lim_{j_M \rightarrow \infty} \sum^{j_M}_{j=0} B_j(\eta^{-1})
\langle m+j | \hat \rho_{meas}| n+j \rangle
\label{inversion}
\end{equation}
with
\begin{equation}
B_j(\eta)=\eta^{(m+n)/2}(1-\eta)^j \left[ {m+j \choose m} 
{n+j \choose n}\right]^{1/2} \, . 
\end{equation}
We have shown \cite{KHL} that the convergence of the series (\ref{inversion})
is not necessarily restricted to the range $0.5 < \eta \leq 1$, although the 
matrix elements $B_j(\eta^{-1})$ of the inverse transformation are divergent 
if $\eta < 0.5$ \cite{interesting}. 
Furthermore \cite{Herzog} 
we could employ multiple runs of compensation or the analytic 
continuation of the series (\ref{inversion}). In this case no 
convergence bound on $\eta$ does exist and hence loss compensation is 
possible {\it in principle}. 

{\it In practice}, of course, experimental imperfections will affect the 
compensation of detection losses \cite{problems}. We have never claimed that
loss compensation is easy \cite{LPjmo}, in fact, we have stressed \cite{KHL} 
that for $\eta < 0.5$ other errors are amplified, e.g. the effect of any
uncertainty in $\eta$ itself.
D'Ariano and Macchiavello \cite{DM} considered the influence of 
statistical errors for homodyne measurements. They assumed a finite number 
$N$ of experiments, i.e. a finite statistical ensemble of individual 
quantum systems. 
In this case the reconstructed density matrix $\langle m | \hat \rho_{meas}
(N) |n \rangle$ is an estimation of the matrix $\langle m | \hat \rho_{meas}
|n \rangle$ with statistical error bars calculated according to Ref. 
\cite{LMKRR}. However, as a fundamental axiom of quantum mechanics, the 
estimation must tend to the ensemble average when the ensemble size 
approaches infinity.
Therefore, if we keep the cut-off $j_M$ in the series (\ref{inversion})
at an arbitrarily large, but fixed value and increase the number $N$ of 
experimental runs, we must approach the correct result $\langle m | \hat 
\rho_{sig} |n \rangle$, with an arbitrarily small, fixed systematic error
that depends only on the cut-off. 
D'Ariano and Macchiavello \cite{DM} did exactly the opposite.
They fixed the size $N$ of the statistical ensemble and increased the cut-off 
$j_M$, and found that the series diverges for $\eta<0.5$. Our Fig 1. 
illustrates the influence of varying both $j_M$ and $N$. We see clearly that 
the order of the limits $j_M \rightarrow \infty$ and $N \rightarrow \infty$ 
is important.
We also see that quite a large number $N$ of runs is required to produce 
faithful data for compensating a low efficiency $\eta$. 
D'Ariano and Macchiavello \cite{DM} 
discussed certainly an interesting aspect of loss compensation but, as 
we have seen, their analysis was incomplete.

Is the compensation bound \cite{DLP} relevant for some more fundamental
features of quantum mechanics than merely technical points of measurement 
technology? Does the $50\%$ bound \cite{DLP} rule out the state measurement 
of an individual system \cite{DY}?
If the bound existed it clearly would, as was pointed out in Ref. \cite{DY}.
From the conclusion, however, does not follow the premise. 
The impossibility of measuring the wave function does not imply that the 
$50 \%$ compensation bound exists.

The problem discussed in Ref. \cite{DY} is again a matter of performing 
limits in the right order. Suppose one taps a series of $N$ 
probe beams from an individual light mode and performs a state reconstruction 
using the $N$ probes as a statistical ensemble. The effect of tapping, i.e.
beam-splitting, is equivalent to detection losses \cite{LP}
with an efficiency $\eta$ scaling like $N^{-1}$.
Therefore, when we attempt to reconstruct the quantum state
of the individual light mode 
we should employ infinitely many probes,
yet with infinitely poor efficiency.
Not surprisingly, we cannot 
compensate the losses in this situation.
A general $50 \%$ efficiency bound is much too much to be required for such a 
delicate matter.

In conclusion, compensation for low overall detection efficiency 
is numerically difficult. 
The value $0.5$ for $\eta$ plays clearly a crucial role \cite{DLP}
because at this value the matrix elements in the 
inverse Bernoulli transformation become unbounded \cite{KHL}. 
Our analysis shows, however, that 0.5 
is neither a rigorous bound for compensating losses 
in optical homodyne tomography, 
nor is this bound required for ruling out the state measurement
of an individual system \cite{DY}.

T.K. acknowledges the support by the National Research Fund of Hungary (OTKA) 
under 
Contract No. F017381 and Hungarian Academy of Sciences Grant No. 96/64-17
and by an E\"otv\"os Fellowship.
T.K. is grateful
to Professor W. P. Schleich for his kind hospitality during his stay in Ulm.
U.L. was supported by the Deutsche Forschungsgemeinschaft.
%%%%%%%%%%%%%%%%%%%%%%%%%%%%%%%%%%%%%%%%%%%%%%%%%%%%%%%%%%%%%%%%%%%%%%

\narrowtext

\section*{Figure Caption}

Plot of the loss--compensated density--matrix element
$\rho_{00}$ with varying cut--off $j_M$ and ensemble size $N$.
We employed the same thermal state with $\bar{n}=2$ as in the
preceding Comment \cite{DM} and used an efficiency $\eta$ of $0.48$.
We performed Monte--Carlo simulations to model a realistic experimental
situation. First, we reconstructed the density matrix 
$\langle m | \hat \rho_{meas} | n \rangle$
from $N$ runs of the computer experiment
using the method developed in Ref. \cite{LMKRR}.
Then we performed the loss compensation (\ref{inversion}) with
varying cut--off $j_M$.
We found that for a given $j_M$ the reconstructed matrix elements
$\rho_{00}$ do approach the actual value of 
$\langle 0 | \hat \rho_{sig}| 0 \rangle = 0.33$ for increasing numbers $N$
of runs, apart from a small systematic error.
On the other hand, if 
we keep the number $N$ of runs constant and increase the cut-off $j_M$
the matrix element diverges \cite{DM},
as can be seen from the behavior of $\rho_{00}$ for $N=10^3$.
Thus, the order of the limits 
$j_M \rightarrow \infty$ and $N \rightarrow \infty$ 
is vital to the loss--compensation procedures \cite{KHL,Herzog}.

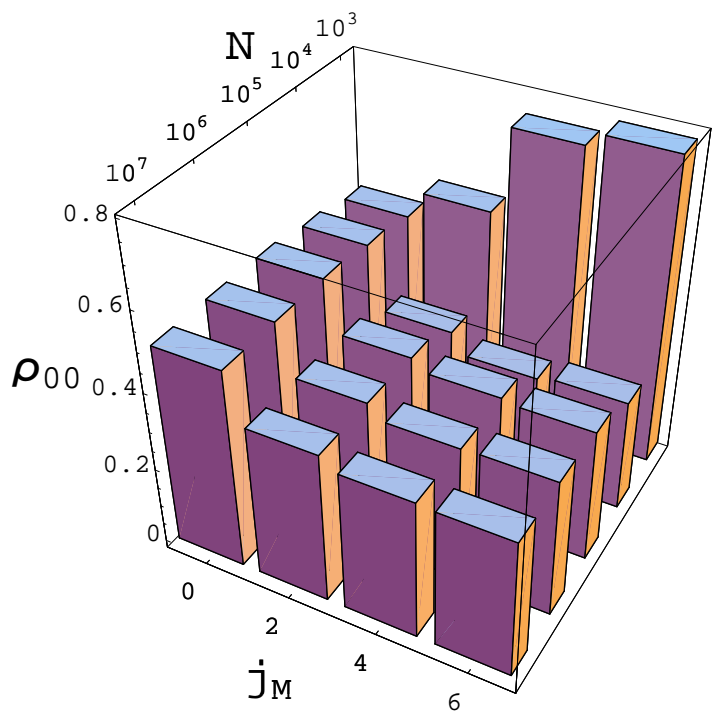
\end{document}